\documentclass[rn,11pt]{ucl_document} %http://mediatools.cs.ucl.ac.uk/nets/dos/browser/transferreport/ucl_document.cls
\pdfoutput=1
\usepackage{graphicx}
\usepackage{hyperref} %/shareopt/tetex/share/texmf/tex/latex/hyperref/hyperref.sty

\begin{document}

\newcommand{\bwa}{{\tt bwa}}

\title{Genetically Improved BarraCUDA}

\date{28 May 2015}
\documentnumber{15/03}

\author{
{W. B.~Langdon}
and
{Brian Yee Hong Lam}
}

\maketitle

\begin{abstract}
BarraCUDA is a C program which
uses the BWA algorithm in parallel with
nVidia CUDA to align short
next generation DNA sequences against a reference genome.
The genetically improved (GI) code is up to three times faster
on short paired end reads from The 1000 Genomes Project
and 60\% more accurate
on a short 
{BioPlanet.com} GCAT alignment benchmark.
GPGPU Barracuda running on a single K80 Tesla GPU
can align short paired end nextgen sequences
up to ten times faster
than \bwa{} on a 12 core CPU.
\end{abstract}

\section{Why Run Bioinformatics on Gaming Machines}

\begin{figure} [!b]
\centerline{\includegraphics{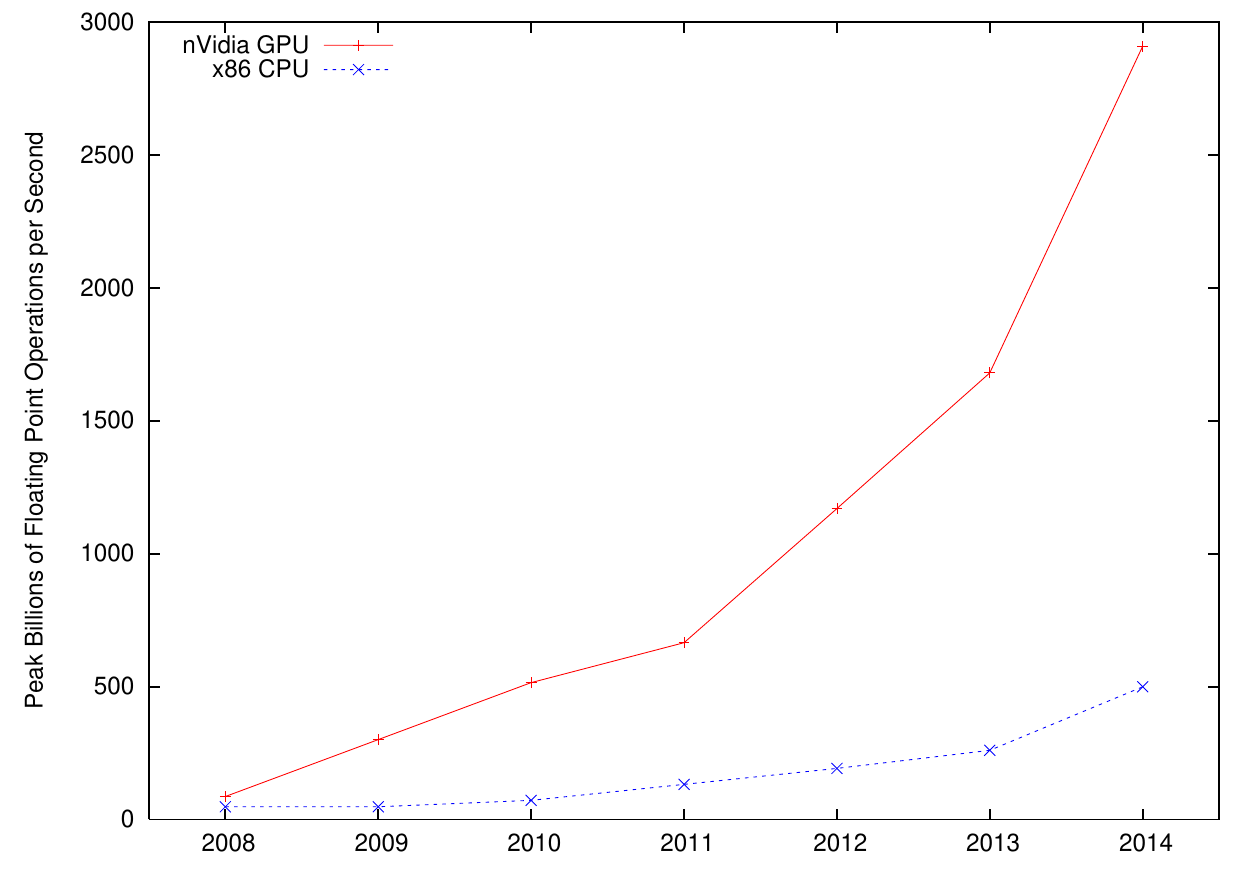}} % gnuplot epstopdf
\caption{
Exponential growth in {\em peak} processing power.
Data from nVidia
\label{fig:GPUvsCPU_data}
}
\end{figure}

The explosive growth in Biological datasets 
has coincided with a similar exponential increase in
computer processing power
(known as Moore's Law~\cite{Moo65}).
Before 2005 the doubling of integrated circuit complexity
every 18 months,
went hand-in-hand with doubling of computer processor clock speeds.
However in the last ten years clock speeds have increased little.
This has not (as yet) limited the exponential growth in 
Bioinformatics datasets and hence processing demand.
Fortunately Moore's Law has continued to apply to the number of
transistors per silicon chip.
Whilst some of these extra transistors have been used to support more
powerful computer instructions,
largely they have been and will continue to be used to support
parallel computing.
In 2005 a typical computer contain one CPU,
nowadays quad code (i.e.\ 4 CPUs) are common place 
with 6, 8 and 12 cores also being available.
This trend will continue.

Modern consumer applications demand high quality and instant response.
With user interfaces containing millions of display elements (pixels) and
thousands of input sensors, the only practical approach has been
parallel processing. Rather than using several CPUs, hardware
dedicated to graphical displays typically contains
hundreds or even thousands of
processing elements. As each each pixel is processed in the same way,
the graphics processing units (GPUs) can take short cuts 
in the hardware. For example, since each of the hundreds of
pixel processing programs is doing exactly the same thing,
the logic to decode program instructions can be shared.
This means
the transistors used to decode the program actually
drive many streaming processing cores (rather than just one).
The research and development of these specialised but highly parallel
graphics accelerator cards
has been paid for largely by the consumer gaming market.
One of the main players in this market is nVidia.
They have sold
hundreds of millions of their GPUs.
(These GPUs are capable of running CUDA,
which is nVidia's general purpose framework for programming their GPUs.
It is used by BarraCUDA.)

About the time of the end of the serial processor clock speed boom,
computer scientists and engineers started to treat GPUs as low cost but
highly parallel computers
and started using their GPUs for general purpose computing
(GPGPUs \cite{owens:2008:ieee}).
This trend continues.
Indeed GPGPU has been combined 
with some enormous volunteer user cloud systems.
For example, much of the raw computer power used by the SETI@HOME
project is actually derived from GPUs within domestic PCs.

Another aspect of GPGPU, has been the introduction
by both nVidia and Intel
of ``screen less'' GPUs,
where the hardware is dedicate to computing applications rather than
computer graphics.
Indeed today half of the ten fastest computers on the planet are based
on GPUs
(\href{http://www.top500.org/}{http://www.top500.org/} May 2015).

Bioinformaticians have not been slow in seizing 
the advantages of GPGPU programming.
CUDA versions of several popular applications have been
written.
However, as with other branches of super computing,
it is often not easy to write code to gain the best of parallel
machines.
For example,
often parallel applications are limited not
by the processing power available but by the time taken to move data
inside the computer to the processing elements.

At the forthcoming GECCO conference \cite{Langdon:2015:GECCO}
we shall present an approach in which a small part of the
manually written code has
been optimised by a variant of %the machine learning algorithm 
genetic programming 
\cite{koza:book,poli08:fieldguide}
to give a huge speed up on that part.
(The raw graphics kernel can process well over a million DNA sequences
a second \mbox{\cite[Fig.~1]{Langdon:2015:GECCO}}.)
The next section will describe the target system,
BarraCUDA~\cite{Klus:2012:BMCrn}.
Section~\ref{sec:details} gives details of the programs 
and DNA benchmarks.
In particular
the standard GCAT Bioinformatics DNA sequence alignment
benchmarks~\cite{Highnam:2015:nc}
and short human paired end next generation DNA sequences taken from 
The 1000 Genomes Project \cite{nature09534}.
This is followed 
(Section~\ref{sec:results}) by 
the overall performance changes
genetic improvement%
~\cite{Langdon:2013:ieeeTEC,Petke:2014:EuroGP,Langdon:2012:mendel,%
jia:2015:gsgp,%
langdon:2015:gi,langdon:2015:hbgpa}
gives
and comparison with \bwa{}.
(See particularly Table~\ref{tab:results}, page~\pageref{tab:results}.)

\section{NextGen DNA Sequence Alignment}

Since the human genome was sequenced in 2000~\cite{ihgsc:2001:nature},
increasingly powerful nextGeneration sequencing machines have
generated vast volumes of short noisy DNA sequences.
Initially sequences where only 30 or so bases long.
(The sequences are stored as strings of the four letters 
A, C, G and T\@. Each character representing one base.)
Where the genetic sequence is variable,
simple statistics 
(i.e., $4^{30} \gg$ length of the genome)
that suggest 30 or so bases would be sufficient to identify where
the sequence lies in the reference genome.
Whilst these data are inevitably noisy,
the main difficulty with this approach is that
(in particular)
the Human genome contains many repeated sequences.
Thus sometimes $4^{30}$ can only identify the repeated pattern
not the location itself.
This lead to 1)~longer sequences but also 
2)~sequencing both ends of much longer sequences.
The second (paired end) approach requires more sophisticated computer
algorithms.
Each end is matched against the reference genome as before.
When an end lies in a repeated sequence,
and so gives multiple match points,
the matches the other end gives are consulted.
As the approximate length of the DNA sequence is known
(or can be inferred),
in many cases potential matches for the two ends can be ignored
as they are simply too far apart.
Paired end analysis is now typical.
Whilst Barracuda can deal both with single ended and paired end DNA
sequences, we shall only benchmark paired end data.

BarraCUDA uses the Burrows-Wheeler algorithm (BWA)
\cite{Bioinformatics-2010-Li-589-95}.
Indeed it source code is derived from a serial implementation
of the BWA algorithm,
simply called \bwa.
Barracuda gets its speed 
by using a GPU to
processing hundreds of thousands of short DNA sequences 
in parallel.
Typically finding where each DNA sequences matches the reference 
human genome is the most time consuming part.
With paired end (pe) data,
Barracuda ``aln'' matches each end separately and then
Barracuda ``sampe'' combines them.
Thus Barracuda sampe does not need a GPU
(although it, unlike \bwa{}~sampe, can exploit multiple CPU cores).

With noise free data and where the DNA sequence matches the 
reference genome exactly,
the Burrows-Wheeler algorithm is relatively straight forward.
Before hand, offline,
the reference genome is encoded into a compressed format
so that all the sequences in the reference genome with the 
same starting subsequences are given the same location in the
compressed file.
Since there are four possible bases, 
extending the prefix sequence by one means this location 
leads to four subsequences prefix strings which are one base longer.
However as the reference genome is finite,
the branching factor quickly falls from four to one.
If a prefix sequence can be followed by exactly one
prefix which is one base longer,
this means all the sequences with this particular prefix have the
same base in the next position.
The index file is arranged to enable rapid sequence look up.
Barracuda and \bwa{} index files are interchangeable.

\pagebreak[4]
On look up, an upper and a lower pointer into the index file are kept.
They span all possible matches,
and so are initially far apart.
As each base in the DNA sequence is processed,
data are read from the index file
and the position of the two pointers are updated.
Actually the distance between them is the number of positions in the
reference genome which match the DNA sequence processed so far.
If the distance becomes one, then there is a unique match.
If the two pointers cross this means the sequence does not exist in
the reference genome.
In good quality data from the 1000 Genomes Project,
about 85\% %give example??
of sequences match uniquely.
Where sequences do not match, this may be either due to noise in the
data or to real mutations in the patient.
To cope with non-exact matches,
the algorithm must carefully back up its search and start 
trying out alternatives.
This slows things down considerably.

For the approach to be feasible, 
Barracuda must load 
the whole of the index file
into the GPU's memory.
Thus the GPU must have enough memory to hold it all.
For the human reference genome,
this means the GPU must have at least four gigabytes of on-board RAM.
Also the 
Burrows-Wheeler algorithm does not allow short cuts.
I.e., every base in the sequence must be processed.
Thus, even before considering mismatches,
Barracuda must make heavy access to the index.
Fortunately modern GPUs have high bandwidth to their on-board memory
(see last column in Table~\ref{tab:gpus}).

\begin{table*}%[tbp]
\caption{\label{tab:gpus}
GPU Hardware.
Year each was announced by nVidia in column~2.
Price (column~3) is either actual (GT~730) or on line quote 
(May~2015,
which may be lower than original list price).
Fourth column is CUDA compute capability level
(as can be used with the nvcc compiler's {\tt -arch} parameter).
Each GPU chip contains 2, 13 or 15 identical 
independent multiprocessors (MP, column~5).
Each MP contains 48 or 192 stream processors (total given in column~7)
whose clock speed is given in column~8.
Onboard memory size and bandwidth are given in the right most two columns.
ECC enabled.
}
\begin{minipage}{\textwidth}
\tabcolsep 4pt
\begin{tabular}{@{}llrcr@{ $\times$ }r@{ $=$ }rrclrr@{}}
GPU &
\multicolumn{3}{r}{compute level} &
\multicolumn{3}{c}{MP \hfill total cores} &
\multicolumn{1}{c}{Clock} & 
\multicolumn{2}{c}{L1/L2 caches} &
\multicolumn{2}{c}{Memory}
\\
\hline
GT 730
\rule[1ex]{0pt}{6pt}
& %\\
2014 &
\pounds 53.89 &
2.1             &
2 & 48 & 96 &%( 2) Multiprocessors, ( 48) CUDA Cores/MP:     96 CUDA Cores
1.40 GHz &%GPU Clock rate:                                1400 MHz (1.40 GHz)
48KB &% 16/32/48KB 
            %https://devtalk.nvidia.com/default/topic/521383/read-only-data-cache-only-for-tesla-or-also-for-gtx-680-/
0.125 MB  &%L2 Cache Size:                                 131072 bytes
4 GB &%Total amount of global memory: 4095 MBytes (4294115328 bytes)
23 GB/s  %Device to Device Bandwidth, PINNED Memory Transfers	23857.3
         %gawk 'END{print 23857.3/1024}' /dev/null 23.2981
\\
Tesla K20
& %\\
2012 &
\pounds 2,905.20 &%Inc VAT
3.5             &%CUDA Capability Major/Minor version number: 3.5
13 & 192 & 2496 &%(13) Multiprocessors x (192) CUDA Cores/MP: 2496 CUDA Cores
0.71 GHz &%GPU Clock rate:                                706 MHz (0.71 GHz)
48KB &% 16/32/48KB 
            %https://devtalk.nvidia.com/default/topic/521383/read-only-data-cache-only-for-tesla-or-also-for-gtx-680-/
1.25 MB  &%L2 Cache Size:                                 1310720 bytes
5 GB &%4.68707 Total amount of global memory: 4800 MBytes (5032706048 bytes)
140 GB/s  %Bandwidth(MB/s) 143386.6
\\
Tesla K40
&%\\
2013 &
\pounds 3,264.83 &
3.5             &%CUDA Capability Major/Minor version number: 3.5
15 & 192 & 2880 &%(13) Multiprocessors x (192) CUDA Cores/MP: 2496 CUDA Cores
0.88 GHz & %GPU Clock rate:                                876 MHz (0.88 GHz)
48KB &% 16/32/48KB 
            %https://devtalk.nvidia.com/default/topic/521383/read-only-data-cache-only-for-tesla-or-also-for-gtx-680-/
1.50 MB  &%L2 Cache Size:                                 1572864 bytes
11 GB %11.2496
&
180 GB/s  %183374.5 Bandwidth(MB/s) gawk 'END{print 183374.5 /1024}'  179.077
\\ 
Tesla K80%
\protect\footnotemark
\protect\footnotetext{K80 is a dual GPU,
performance figures given for one half.}
&%\\
2014 &
\pounds 6,260.65 &
3.7             &%CUDA Capability Major/Minor version number: 3.7
13 & 192 & 2496 &%(13) Multiprocessors, (192) CUDA Cores/MP: 2496 CUDA Cores
0.82 GHz & %GPU Clock rate:                                824 MHz (0.82 GHz)
48KB &% 16/32/48KB 
            %https://devtalk.nvidia.com/default/topic/521383/read-only-data-cache-only-for-tesla-or-also-for-gtx-680-/
1.50 MB  &%L2 Cache Size:                                 1572864 bytes
11 GB %11.2496
&
138 GB/s  %gawk 'END{print 141242.4 /1024}' 137.932
\\ 
\end{tabular}
\end{minipage}
\end{table*}

\begin{table}%[tbp]
\caption{\label{tab:cpus}
CPUs.
The desktop computer houses one GT 730.
The servers are part of the 
Darwin Supercomputer of the University of Cambridge
and hold multiple Tesla K20 or K80 GPUs.
}
\begin{center}
\begin{tabular}{@{}lrrr@{}}
\multicolumn{1}{@{}c}{Type} & 
\multicolumn{1}{c}{Cores} & 
\multicolumn{1}{c}{Clock} & 
\multicolumn{1}{c@{}}{Memory}
\\
\hline
Desktop & 
2 & 2.66 GHz & 4 GB \\ %MemTotal:      3980024 kB 3886.74
Darwin &
12 & 2.60 GHz & 62 GB
\\
NVK80 &
24 & 2.30 GHz & 125 GB \\
\end{tabular}
\end{center}
\end{table}

Typically the Burrows-Wheeler algorithm scales linearly with the 
length of the DNA sequences to be looked up.
This makes it more suitable for shorter sequences
than for longer ones.

Taking The 1000 Genomes Project as an example,
\cite[Fig.~4]{Langdon:2013:BDM}
shows some sequence lengths are much more common than others.
In Section~\ref{sec:results}
we report tests on paired end data 
comprised of 36 bases per end and of 100 bases per end.
Both are common in The 1000 Genomes Project.
In fact the most popular is 101 bases,
which is almost the same as one of the benchmarks provided by
\href{http://www.bioplanet.com}
{BioPlanet}'s
Genome Comparison and Analytic Testing (GCAT) platform
\cite{Highnam:2015:nc}.

\section{Programs,
DNA sequences %1000 Genomes and GCAT Benchmarks 
and
Parallel Operation under Multi-core Unix}
\label{sec:details}

\subsection{\bwa{} 0.7.12} %./bwa
The current release of \bwa{} (May 2015, Version: 0.7.12-r1039),
i.e.\
\href{https://github.com/lh3/bwa/archive/0.7.12.tar.gz}
{bwa-0.7.12.tar.gz},
was down loaded from GitHub
and compiled with default settings %-O2 -DHAVE_PTHREAD -DUSE_MALLOC_WRAPPERS
(i.e.\ including support for multi-threading).

\subsection{Barracuda 0.6.2}  %0.6.2t.exe
For comparison, the previous version of BarraCUDA,
i.e.~0.6.2,
was 
compiled with default settings %-O2 -DHAVE_PTHREAD -DUSE_MALLOC_WRAPPERS
(i.e.~again including support for multi-threading).

\subsection{Barracuda 0.7.107}

The current release (May 2015, Version 0.7.0r107),
bwa-0.7.12.tar.gz,
was down loaded from 
\href{http://sourceforge.net/projects/seqbarracuda/files/latest/download}
{SourceForge}.
Again it was built with default setting
(including support for multi-threading).
However a second version was built specifically for the GT~730
which was compiled with
\verb'-arch 2.1' 
to support compute level 2.1,
cf.\ column~4 in Table~\ref{tab:gpus}.
(The default is now compute level 3.5 or higher).

\subsection{Reference Genome: UCSC HG19 ucsc.hg19.fasta.gz}
\label{sec:hg19}

Although GCAT provides a pointer 
(\href{http://hgdownload.cse.ucsc.edu/downloads.html#human}
{http://hgdownload.cse.ucsc.edu/downloads.html\#human})
to UCSC,
the reference human genome was downloaded from the
Broad Institute's GATK resource bundle 
via 
\href{ftp://gsapubftp-anonymous@ftp.broadinstitute.org/bundle/2.8/hg19/}
{FTP}
(approximately
900 %gawk 'END{print 948729977/1024/1024}' /dev/null 904.779
megabytes compressed).
It was converted into two indexes.
Barracuda 0.6.2 converted ucsc.hg19.fasta.gz into an index for itself
(%
4.4~GB)\@. %4705754674 4.38258
Secondly
Barracuda 0.7.0 converted it into an index for itself and for \bwa{}
(5.1~GB). %5490044980 5.113 hg19.sa

\subsection{36 base pairs: 1000 Genomes project} 
\label{sec:fastq_1000}
The 1000 Genomes Project~\cite{nature09534}
has made available a vast volume of data via its 
\href{ftp://ftp.1000genomes.ebi.ac.uk/vol1/ftp/}
{FTP} site.
One of its normal (i.e.~not color space encoded)
paired end data with 36 DNA bases per end
was chosen at random
(ERR001270)
and downloaded in compressed form using wget.
ERR001270 consists of two files
(one per end of the DNA sequence)
each containing
1.1 gigabytes (compressed).
ERR001270 contains
14\,102\,867 %gawk 'END{print 112822936 /8}' /dev/null 14102867
36 base DNA sequences.
Approximately
5.7\% %gawk 'END{print 100*(14102867*2  -  26609021)/(14102867*2)}' /dev/null 5.66095
of sequences occur more than once in ERR001270.
The files are in ``fastq'' format and so
also contain quality values
but these are not used by \bwa{} or by either version of Barracuda.

Initially \bwa{} objected to the sequence names provided
by The 1000 Genome Project.
However this was readily resolved
so that each pair of sequences had its own unique name.

\subsection{100 base pairs: GCAT Benchmark}
\label{sec:fastq_gcat}

BioPlanet.com hosts several 
sequence alignment and variant calling 
next generation DNA benchmarks on its 
\href{http://www.bioplanet.com/gcat}
{GCAT} web pages~\cite{Highnam:2015:nc}.
We report results on their 
100bp-pe-small-indel alignment benchmark (gcat\_set\_037).
This consists of two files (one per end) each of
3~gigabytes %ls -ltr | gawk '{print f+=$5,f/1024/1024/1024}' 3320753270 3.09269
(uncompressed),
each containing  
5\,972\,625 %gawk 'END{print 23890500/4}' /dev/null 5972625
100 base sequences.
(Less than 0.1\% of sequences were repeated.)
The files are again in ``fastq'' format 
but only 
contain dummy quality values 
however again these are not used by \bwa{} or by Barracuda.

\subsection{Parallel operation on multi-core CPUs with bash pipes}
Barracuda and \bwa{} have similar operations and command lines.
For paired end data, 
``aln'' is run separately for each fastq sequence file
(Sections~\ref{sec:fastq_1000} and~\ref{sec:fastq_gcat} above).
For every fastq sequence,
``aln'' produces zero or more possible alignments in the 
reference genome (Section~\ref{sec:hg19})
and saves them in a binary .sai file.
(.sai files are generally not compatible between different versions).
For each pair of sequences,
``sampe'' takes the alignments from the .sai files,
the sequences themselves and the index file for the reference genome
to produce alignments for each end in sam format.

\begin{figure}
\includegraphics[width=\textwidth]{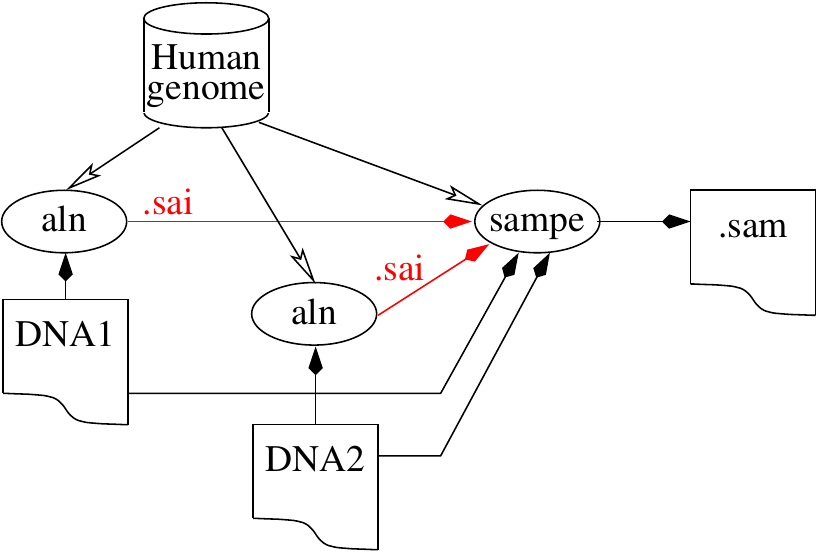} %epstopdf sampe_pipes.eps
\caption{\label{fig:sampe_pipes}
Processing paired end DNA sequences.
``aln'' is run twice (once per end)
and its alignments are piped (red arrows) into ``sampe''
(sam~(pe) paired end).
``sampe'' also reads the index of the reference human genome
and both ends of each DNA sequence in order to give the
combined alignment in sam format.
In the case of barracuda,
the two ``aln'' process each use a GPU
and ``sampe'' uses multiple host threads.
For \bwa{} ``aln'' uses multiple host threads
but ``sampe'' is single threaded.
}
\end{figure}

Notice the .sai files are intermediate and can be deleted after
the .sam file has been created.
However the .sai files are large.
(E.g.\ 
830 mega bytes %830.759
for Barracuda 0.6.2 on the 1000 Genomes Project example
and
340 mega bytes for the GCAT example.)
Since the .sai files are both written to and read sequentially,
under Unix using bash,
it is not necessary to explicitly store them.
Instead
``aln'' can write them to a unix pipe
and ``sampe'' can read them from those pipes
(see Figures~\ref{fig:sampe_pipes} and~\ref{fig:bash}).
With large memory multi-core CPUs it is quite feasible to
run both ``aln'' processes and the ``sampe'' process in parallel
(see Table~\ref{tab:cpus}).

\begin{figure} %[!b]
\begin{minipage}{\textwidth}
\begin{verbatim}
  $exe1 sampe -t 24 $hg19			\
      <($exe1 aln -C 0 $hg19 $seq1)		\
      <($exe1 aln -C 1 $hg19 $seq2) $seq1 $seq2	\
    > $sam
\end{verbatim}
\end{minipage}
\caption{\label{fig:bash}
Example bash command line using process substitution, pipes and 
input-output redirection to run two ``aln'' processes in parallel with
``sampe'', thus avoiding use of intermediate disk files.
{\tt 
\$exe1,
\$hg19,
\$seq1,
\$seq2} and
{\tt \$sam}
are the names of bash environment variables.
{\tt\$exe1} is the program,
{\tt\$hg19} the location of the reference genome index,
{\tt\$seq1} and {\tt\$seq2} are the files holding the pairs of DNA sequences
and {\tt\$sam} is the output.
See also Figure~\protect\ref{fig:sampe_pipes}.
}
\end{figure}

The sam files are plain text and also large,
nearly 6~GB for ERR001270
and
well over 4~GB for the GCAT benchmark.
GCAT uses the compressed binary format bam,
even so
gcat\_set\_037.bam is almost a gigabyte.
gcat\_set\_037.sam was converted to gcat\_set\_037.bam by samtools.
The time for this post processing 
(a few minutes)
is not included in Table~\ref{tab:results}.

\vspace*{-1ex}
\subsection{Problems and Work Arounds}
\label{sec:bugs}
\vspace*{-2ex}

A single GeForce GT 730 was available. 
It was mounted in a desktop linux PC with 4~GB of RAM\@.
This was enough to run ``aln'' but not ``sampe''.
(The .sai files could be transferred to a much larger
linux server to run ``sampe''.)
Typically in Barracuda ``sampe'' takes little time
and on a typical large multi-core server
can be run in parallel with the two ``aln'' processes
with little impact on total wall clock time
(see Figures~\ref{fig:sampe_pipes} and~\ref{fig:bash}).
For ease of comparison
the data in Table~\ref{tab:results} are an estimate 
for two GT~730's mounted in a large multi-core CPU\@.
They are calculated from the wall clock time of the 
slowest of the two .sai files.

Barracuda version 0.6.2 beta
``sampe'' failed on ERR001270
if run with multiple threads.
The only work around was to used \verb'-t 1' to prevent 
use of multiple threads.
(Bug fixed before version 0.7.107.)

There is a small HTML web page which lists a few issues 
(some open but several now fixed)
and common misunderstandings
about Barracuda
at 
\href{http://www.cs.ucl.ac.uk/staff/W.Langdon/barracuda/}
{http://www.cs.ucl.ac.uk/staff/W.Langdon/barracuda/}.

\section{Results}
\label{sec:results}

\bwa{} and the original and the GI improved version of Barracuda
were each run five times on both 
the fourteen million real world paired end DNA sequences
from The 1000 Genomes Project
(Section~\ref{sec:fastq_1000})
and the almost six million paired end DNA sequences
provided by GCAT as a benchmark
(Section~\ref{sec:fastq_gcat}).
\bwa{} was run on 12 core 2.60~GHz CPUs
(see Table~\ref{tab:cpus})
whilst Barracuda was run on three GPUs,
stretching from \pounds 50 low end GT~730
to the top of the range K80 Tesla
(see Table~\ref{tab:gpus}).
The results are summarised in Table~\ref{tab:results}.

Apart from the low end GT~730,
Barracuda is typically between two and ten times faster
than
the current release of
\bwa{} on a 12~core CPU
(see
figures within round brackets in the upper part of 
Table~\ref{tab:results}).
The lower part of Table~\ref{tab:results}
presents the Barracuda data in the upper part as ratios between the
previous release of Barracuda (0.6.2) with the current
genetically improved~\cite{Langdon:2015:GECCO}
version (0.7.107).
Table~\ref{tab:results} shows the newer version is up to three times
faster on the real world DNA sequences (36bp) and 
typically about 10\% faster on the longer benchmark strings (100bp).

\begin{table}%[tbp]
\caption{\label{tab:results}
Mean number of paired end sequences processed per second.
In (brackets) speed relative to \bwa{}~0.7.12.
$\pm$ gives standard deviation estimated from five runs.
There was almost no variation in accuracy reported by GCAT\@.
}
\begin{minipage}{\textwidth}
\tabcolsep 4pt
\begin{tabular}{@{}lrr@{ $\pm$}l@{}*{3}{r@{ $\pm$}r@{ }r@{ $\pm$}l}c@{}}
Prog & length & 
\multicolumn{6}{c}{12 core CPU\protect\footnote{2.60GHz,
see ``Darwin'' in Table~\protect\ref{tab:cpus}}\hfill %} &
GT 730%
\protect\footnote{Estimated for two GT 730 GPUs. 
See Section~\protect\ref{sec:bugs}}\hfill\hfill} &
\multicolumn{4}{c}{2$\times$ K20} &
\multicolumn{5}{c@{}}{\hfill\hfill\hfill K80 \hfill%} &
GCAT Accuracy\%}
\\\hline
%bwa_speed.bat Revision: 1.11 crest29.cs.ucl.ac.uk
\bwa & 36bp &
%bwa_speed_1.bat $Revision: 1.3 $ crest29.cs.ucl.ac.uk /cs/research/crest/gpu/ucacbbl/barracuda_work/gcat
%bwa_speed.awk Revision: 1.9  N=14102867 25 May 2015
%elapse_gcat.awk Revision: 1.7  
%1431518345 8 threads Wed May 13 12:59:05 BST
%7443 slurm-1331198.out #bwa-0.7.12/bwa-0.7.12/bwa sam and pipes done Wed May 13 15:03:08 BST 2015
%elapse_gcat.awk Revision: 1.7  
%1431539634 8 threads Wed May 13 18:53:54 BST
%7388 slurm-1331687.out #bwa-0.7.12/bwa-0.7.12/bwa sam and pipes done Wed May 13 20:57:02 BST 2015
%elapse_gcat.awk Revision: 1.7  
%1431551063 8 threads Wed May 13 22:04:23 BST
%7208 slurm-1332202.out #bwa-0.7.12/bwa-0.7.12/bwa sam and pipes done Thu May 14 00:04:31 BST 2015
%elapse_gcat.awk Revision: 1.7  
%1431588267 8 threads Thu May 14 08:24:27 BST
%7761 slurm-1332549.out #bwa-0.7.12/bwa-0.7.12/bwa sam and pipes done Thu May 14 10:33:48 BST 2015
%elapse_gcat.awk Revision: 1.7  
%1431597183 8 threads Thu May 14 10:53:03 BST
%7476 slurm-1332656.out #bwa-0.7.12/bwa-0.7.12/bwa sam and pipes done Thu May 14 12:57:39 BST 2015
%7455.2 & 199.824 %278059754
 1900 &  50
&
\multicolumn{4}{c}{-} &
\multicolumn{4}{c}{-} &
\multicolumn{4}{c}{-} &
\multicolumn{1}{c}{-} \\
\bwa & 100bp &
%bwa_speed_1.bat $Revision: 1.3 $ crest29.cs.ucl.ac.uk /cs/research/crest/gpu/ucacbbl/barracuda_work/gcat
%bwa_speed.awk Revision: 1.9  N=5972625 25 May 2015
%elapse_gcat.awk Revision: 1.7  
%1430937602 8 threads Wed May 6 19:40:02 BST
%1331 slurm-1318728.out #bwa sam and pipes done Wed May 6 20:02:13 BST 2015
%elapse_gcat.awk Revision: 1.7  
%1430940061 8 threads Wed May 6 20:21:01 BST
%1331 slurm-1318760.out #bwa sam and pipes done Wed May 6 20:43:12 BST 2015
%elapse_gcat.awk Revision: 1.7  
%1430941682 8 threads Wed May 6 20:48:02 BST
%1328 slurm-1318771.out #bwa sam and pipes done Wed May 6 21:10:10 BST 2015
%elapse_gcat.awk Revision: 1.7  
%1430982608 8 threads Thu May 7 08:10:08 BST
%1320 slurm-1319044.out #bwa sam and pipes done Thu May 7 08:32:08 BST 2015
%elapse_gcat.awk Revision: 1.7  
%1430986330 8 threads Thu May 7 09:12:10 BST
%1319 slurm-1319062.out #bwa sam and pipes done Thu May 7 09:34:09 BST 2015
%1325.8 & 5.89067 %8788867
 4500 &  20
&
\multicolumn{4}{c}{-} &
\multicolumn{4}{c}{-} &
\multicolumn{4}{c}{-} &
98.91 \\
0.6.2 & 36bp &
\multicolumn{2}{c}{-} &
 3270 &   2 & (1.7 & 0.05) %(1.72846 0.0463488)
&
%bwa_speed_1.bat $Revision: 1.3 $ crest29.cs.ucl.ac.uk /cs/research/crest/gpu/ucacbbl/barracuda_work/gcat
%bwa_speed.awk Revision: 1.9  N=14102867 25 May 2015
%elapse_gcat.awk Revision: 1.7  
%1431674059 Using pipes Fri May 15 08:14:19 BST
%2680 slurm-1333776.out #./0.6.2t.exe sam and pipes done Fri May 15 08:58:59 BST 2015
%elapse_gcat.awk Revision: 1.7  
%1431679953 Using pipes Fri May 15 09:52:33 BST
%2640 slurm-1333831.out #./0.6.2t.exe sam and pipes done Fri May 15 10:36:33 BST 2015
%elapse_gcat.awk Revision: 1.7  
%1431685522 Using pipes Fri May 15 11:25:22 BST
%2722 slurm-1333887.out #./0.6.2t.exe sam and pipes done Fri May 15 12:10:44 BST 2015
%elapse_gcat.awk Revision: 1.7  
%1431694016 Using pipes Fri May 15 13:46:56 BST
%2567 slurm-1333979.out #./0.6.2t.exe sam and pipes done Fri May 15 14:29:43 BST 2015
%elapse_gcat.awk Revision: 1.7  
%1431697447 Using pipes Fri May 15 14:44:07 BST
%2653 slurm-1334388.out #./0.6.2t.exe sam and pipes done Fri May 15 15:28:20 BST 2015
%2652.4 & 57.1253 %35189182
 5300 & 110 & (2.8 & 0.10) %(2.81074 0.0966448)
&
%bwa_speed_1.bat $Revision: 1.3 $ crest29.cs.ucl.ac.uk /cs/research/crest/gpu/ucacbbl/barracuda_work/gcat
%bwa_speed.awk Revision: 1.9  N=14102867 25 May 2015
%elapse_gcat.awk Revision: 1.7  
%1432495751 Using pipes Sun May 24 20:29:11 BST
%2231 slurm-1352831.out #./0.6.2t.exe sam and pipes done Sun May 24 21:06:22 BST 2015
%elapse_gcat.awk Revision: 1.7  
%1432498288 Using pipes Sun May 24 21:11:28 BST
%2111 slurm-1352843.out #./0.6.2t.exe sam and pipes done Sun May 24 21:46:39 BST 2015
%elapse_gcat.awk Revision: 1.7  
%1432500819 Using pipes Sun May 24 21:53:39 BST
%2193 slurm-1352853.out #./0.6.2t.exe sam and pipes done Sun May 24 22:30:12 BST 2015
%elapse_gcat.awk Revision: 1.7  
%1432534229 Using pipes Mon May 25 07:10:29 BST
%2123 slurm-1354020.out #./0.6.2t.exe sam and pipes done Mon May 25 07:45:52 BST 2015
%elapse_gcat.awk Revision: 1.7  
%1432536723 Using pipes Mon May 25 07:52:03 BST
%2242 slurm-1354030.out #./0.6.2t.exe sam and pipes done Mon May 25 08:29:25 BST 2015
%2180 & 60.4649 %23776624
 6500 & 180 & (3.4 & 0.13) %(3.41982 0.131905)
&
\multicolumn{1}{c}{-} \\
0.6.2 & 100bp &
\multicolumn{2}{c}{-} &
 1860 &   4 & (0.4 & 0.002) %(0.412829 0.00212275)
&
%bwa_speed_1.bat $Revision: 1.3 $ crest29.cs.ucl.ac.uk /cs/research/crest/gpu/ucacbbl/barracuda_work/gcat
%bwa_speed.awk Revision: 1.9  N=5972625 25 May 2015
%elapse_gcat.awk Revision: 1.7  
%1431355636 Using pipes Mon May 11 15:47:16 BST
%681 slurm-1327869.out #./0.6.2t.exe sam and pipes done Mon May 11 15:58:37 BST 2015
%elapse_gcat.awk Revision: 1.7  
%1431356772 Using pipes Mon May 11 16:06:12 BST
%668 slurm-1327878.out #./0.6.2t.exe sam and pipes done Mon May 11 16:17:20 BST 2015
%elapse_gcat.awk Revision: 1.7  
%1431358672 Using pipes Mon May 11 16:37:52 BST
%681 slurm-1327916.out #./0.6.2t.exe sam and pipes done Mon May 11 16:49:13 BST 2015
%elapse_gcat.awk Revision: 1.7  
%1431359471 Using pipes Mon May 11 16:51:11 BST
%698 slurm-1327950.out #./0.6.2t.exe sam and pipes done Mon May 11 17:02:49 BST 2015
%elapse_gcat.awk Revision: 1.7  
%1431360237 Using pipes Mon May 11 17:03:57 BST
%686 slurm-1328062.out #./0.6.2t.exe sam and pipes done Mon May 11 17:15:23 BST 2015
%682.8 & 10.8028 %2331546
 8700 & 140 & (1.9 & 0.03) %(1.94171 0.0319088)
&
%bwa_speed_1.bat $Revision: 1.3 $ crest29.cs.ucl.ac.uk /cs/research/crest/gpu/ucacbbl/barracuda_work/gcat
%bwa_speed.awk Revision: 1.9  N=5972625 25 May 2015
%elapse_gcat.awk Revision: 1.7  
%1431017460 Using pipes Thu May 7 17:51:00 BST
%516 slurm-1320160.out #./0.6.2t.exe sam and pipes done Thu May 7 17:59:36 BST 2015
%elapse_gcat.awk Revision: 1.7  
%1431018806 Using pipes Thu May 7 18:13:26 BST
%509 slurm-1320176.out #./0.6.2t.exe sam and pipes done Thu May 7 18:21:55 BST 2015
%elapse_gcat.awk Revision: 1.7  
%1431021285 Using pipes Thu May 7 18:54:45 BST
%507 slurm-1320199.out #./0.6.2t.exe sam and pipes done Thu May 7 19:03:12 BST 2015
%elapse_gcat.awk Revision: 1.7  
%1431022570 Using pipes Thu May 7 19:16:10 BST
%511 slurm-1320222.out #./0.6.2t.exe sam and pipes done Thu May 7 19:24:41 BST 2015
%elapse_gcat.awk Revision: 1.7  
%1431023148 Using pipes Thu May 7 19:25:48 BST
%505 slurm-1320227.out #./0.6.2t.exe sam and pipes done Thu May 7 19:34:13 BST 2015
%509.6 & 4.219 %1298532
11700 & 100 & (2.6 & 0.02) %(2.60165 0.024445)
&
97.49 \\
0.7.107 & 36bp &
\multicolumn{2}{c}{-} &
 7600 &   6 & (4.0 & 0.11) %(4.01591 0.107688)
&
%bwa_speed_1.bat $Revision: 1.3 $ crest29.cs.ucl.ac.uk /cs/research/crest/gpu/ucacbbl/barracuda_work/gcat
%bwa_speed.awk Revision: 1.9  N=14102867 25 May 2015
%elapse_gcat.awk Revision: 1.7  
%1431446907 Using pipes Tue May 12 17:08:27 BST
%1106 slurm-1329936.out #./barracuda_0.7.107.exe sam and pipes done Tue May 12 17:26:53 BST 2015
%elapse_gcat.awk Revision: 1.7  
%1431448214 Using pipes Tue May 12 17:30:14 BST
%1101 slurm-1330016.out #./barracuda_0.7.107.exe sam and pipes done Tue May 12 17:48:35 BST 2015
%elapse_gcat.awk Revision: 1.7  
%1431449454 Using pipes Tue May 12 17:50:54 BST
%1091 slurm-1330034.out #./barracuda_0.7.107.exe sam and pipes done Tue May 12 18:09:05 BST 2015
%elapse_gcat.awk Revision: 1.7  
%1431457479 Using pipes Tue May 12 20:04:39 BST
%1074 slurm-1330331.out #./barracuda_0.7.107.exe sam and pipes done Tue May 12 20:22:33 BST 2015
%elapse_gcat.awk Revision: 1.7  
%1431504318 Using pipes Wed May 13 09:05:18 BST
%1079 slurm-1330623.out #./barracuda_0.7.107.exe sam and pipes done Wed May 13 09:23:17 BST 2015
%1090.2 & 13.7368 %5943435
12900 & 160 & (6.8 & 0.20) %(6.83838 0.202534)
&
%bwa_speed_1.bat $Revision: 1.3 $ crest29.cs.ucl.ac.uk /cs/research/crest/gpu/ucacbbl/barracuda_work/gcat
%bwa_speed.awk Revision: 1.9  N=14102867 25 May 2015
%elapse_gcat.awk Revision: 1.7  
%1431610827 Using pipes Thu May 14 14:40:27 BST
%736 slurm-1332951.out #./barracuda_0.7.107.exe sam and pipes done Thu May 14 14:52:43 BST 2015
%elapse_gcat.awk Revision: 1.7  
%1431613265 Using pipes Thu May 14 15:21:05 BST
%716 slurm-1333000.out #./barracuda_0.7.107.exe sam and pipes done Thu May 14 15:33:01 BST 2015
%elapse_gcat.awk Revision: 1.7  
%1431614057 Using pipes Thu May 14 15:34:17 BST
%690 slurm-1333029.out #./barracuda_0.7.107.exe sam and pipes done Thu May 14 15:45:47 BST 2015
%elapse_gcat.awk Revision: 1.7  
%1431615202 Using pipes Thu May 14 15:53:22 BST
%698 slurm-1333049.out #./barracuda_0.7.107.exe sam and pipes done Thu May 14 16:05:00 BST 2015
%elapse_gcat.awk Revision: 1.7  
%1431619536 Using pipes Thu May 14 17:05:36 BST
%707 slurm-1333159.out #./barracuda_0.7.107.exe sam and pipes done Thu May 14 17:17:23 BST 2015
%709.4 & 17.7708 %2517505
19900 & 500 & (10.5 & 0.39) %(10.5092 0.38555)
&
\multicolumn{1}{c}{-} \\
0.7.107 & 100bp &
\multicolumn{2}{c}{-} &
 2100 &  14 & (0.5 & 0.004) %(0.467086 0.0037817)
&
%bwa_speed_1.bat $Revision: 1.3 $ crest29.cs.ucl.ac.uk /cs/research/crest/gpu/ucacbbl/barracuda_work/gcat
%bwa_speed.awk Revision: 1.9  N=5972625 25 May 2015
%elapse_gcat.awk Revision: 1.7  
%1431077063 Using pipes Fri May 8 10:24:23 BST
%683 slurm-1320667.out #./barracuda_0.7.107.exe sam and pipes done Fri May 8 10:35:46 BST 2015
%elapse_gcat.awk Revision: 1.7  
%1431078601 Using pipes Fri May 8 10:50:01 BST
%675 slurm-1320684.out #./barracuda_0.7.107.exe sam and pipes done Fri May 8 11:01:16 BST 2015
%elapse_gcat.awk Revision: 1.7  
%1431079382 Using pipes Fri May 8 11:03:02 BST
%674 slurm-1320692.out #./barracuda_0.7.107.exe sam and pipes done Fri May 8 11:14:16 BST 2015
%elapse_gcat.awk Revision: 1.7  
%1431080275 Using pipes Fri May 8 11:17:55 BST
%680 slurm-1320699.out #./barracuda_0.7.107.exe sam and pipes done Fri May 8 11:29:15 BST 2015
%elapse_gcat.awk Revision: 1.7  
%1431081156 Using pipes Fri May 8 11:32:36 BST
%686 slurm-1320781.out #./barracuda_0.7.107.exe sam and pipes done Fri May 8 11:44:02 BST 2015
%679.6 & 5.12835 %2309386
 8800 &  70 & (2.0 & 0.02) %(1.95085 0.0170837)
&
%bwa_speed_1.bat $Revision: 1.3 $ crest29.cs.ucl.ac.uk /cs/research/crest/gpu/ucacbbl/barracuda_work/gcat
%bwa_speed.awk Revision: 1.9  N=5972625 25 May 2015
%elapse_gcat.awk Revision: 1.7  
%1431008951 Using pipes Thu May 7 15:29:11 BST
%482 slurm-1319955.out #./barracuda_0.7.107.exe sam and pipes done Thu May 7 15:37:13 BST 2015
%elapse_gcat.awk Revision: 1.7  
%1431009783 Using pipes Thu May 7 15:43:03 BST
%464 slurm-1319968.out #./barracuda_0.7.107.exe sam and pipes done Thu May 7 15:50:47 BST 2015
%elapse_gcat.awk Revision: 1.7  
%1431010795 Using pipes Thu May 7 15:59:55 BST
%462 slurm-1320019.out #./barracuda_0.7.107.exe sam and pipes done Thu May 7 16:07:37 BST 2015
%elapse_gcat.awk Revision: 1.7  
%1431011450 Using pipes Thu May 7 16:10:50 BST
%474 slurm-1320031.out #./barracuda_0.7.107.exe sam and pipes done Thu May 7 16:18:44 BST 2015
%elapse_gcat.awk Revision: 1.7  
%1431012091 Using pipes Thu May 7 16:21:31 BST
%458 slurm-1320041.out #./barracuda_0.7.107.exe sam and pipes done Thu May 7 16:29:09 BST 2015
%468 & 9.79796 %1095504
12800 & 270 & (2.8 & 0.06) %(2.83291 0.0606301)
&
98.43 \\
\hline\hline
\multicolumn{17}{c}{\rule[-1ex]{0pt}{4ex}Improvement ratio Barracuda 0.7.107 over 0.6.2}
\\
& 36bp &
\multicolumn{2}{c}{-} &
%bwa_speed.awk Revision: 1.9  N= 25 May 2015
\multicolumn{4}{c}{2.32 $\pm$ 0.003}% 2.32341  0.00262327
&
%bwa_speed.awk Revision: 1.9  N= 25 May 2015
\multicolumn{4}{c}{2.43 $\pm$ 0.06\rule{1ex}{0pt}}% 2.43295  0.0607077
&
%bwa_speed.awk Revision: 1.9  N= 25 May 2015
\multicolumn{4}{c}{3.07 $\pm$ 0.11\rule{1ex}{0pt}}% 3.07302  0.114851
&
\multicolumn{1}{c}{-}
\\
& 100bp &
\multicolumn{2}{c}{-} &
%bwa_speed.awk Revision: 1.9  N= 25 May 2015
\multicolumn{4}{c}{1.13 $\pm$ 0.01\rule{1ex}{0pt}}% 1.13143  0.00819866
&
%bwa_speed.awk Revision: 1.9  N= 25 May 2015
\multicolumn{4}{c}{1.00 $\pm$ 0.02\rule{1ex}{0pt}}% 1.00471  0.0176113
&
%bwa_speed.awk Revision: 1.9  N= 25 May 2015
\multicolumn{4}{c}{1.09 $\pm$ 0.02\rule{1ex}{0pt}}% 1.08889  0.0245145
&
1.60
\\

\end{tabular}
\end{minipage}
\end{table}

\pagebreak[4]
\section{Discussion}

\bwa{} ``sampe'' %(unlike \bwa{} ``aln'')
does not support multiple host threads.
As shorter sequences are expected to give rise to more duplicate
matches and so give ``sampe'' more work to do,
this may explain why \bwa{} performs relatively badly
on the short 36~bp 1000 Genomes Project data %.
(see row \bwa{} 36bp in Table~\ref{tab:results}).
On the 1000 Genomes Project example (36bp), 
even a lowly GT~730
running Barracuda~0.7.107
can beat \bwa{}
on at 12 core super computer node.
On the longer GCAT benchmark (100bp),
{\em one} GT~730 is about quarter of the speed of the 12 core computer.
However both the pair of K20s and the K80 are faster than \bwa{}
on both real world and GCAT examples.

On the GCAT benchmark,
the new version of Barracuda is more accurate than 0.6.2
and approaches the accuracy of \bwa{}~0.7.12.

The variability of run time on the super computer
(Table~\ref{tab:results} columns~4, 10 and~14)
is typical of use on shared disk systems.
In contrast, typically elapse times of CUDA kernels are very consistent
(cf.\ Table~\ref{tab:results} column~6).
We anticipate slightly higher and 
but more stable performance could be achieved
on the super computer by placing files on local or ram disks.
(Perhaps a couple of percent improvement may be obtained
by using a ramdisk.)
However ignoring the time to transfer data files 
within the super computer might give unrepresentative results.

\section{Conclusions}

The new version of Barracuda, 
particularly on large real world examples,
is a substantial improvement.
Details of the genetic improvement process used to both tune its parameters and
code will shortly be presented at the GECCO conference
\cite{Langdon:2015:GECCO}.
Both the new version and the genetic improvement process
are freely available.
(The genetically improved version of BarraCUDA 
has been in use via 
\href{http://sourceforge.net/projects/seqbarracuda/?source=typ_redirect}
{SourceForge} since 20 March 2015.
The GI code may be down loaded from the author's 
FTP site from file
\href{http://www.cs.ucl.ac.uk/staff/W.Langdon/ftp/gp-code/barracuda_gp.tar.gz}
{gp-code/barracuda\_gp.tar.gz}.)

Depending upon examples,
even a \pounds 50 GPU running Barracuda can be faster than \bwa{}
on a twelve core 2.60~GHz CPU\@.
With a top end nVidia Tesla GPU,
Barracuda can be more than ten times faster
than \bwa{} on a 12~core CPU\@.

\subsection*{Acknowledgements}

\noindent
I am grateful for the assistance of
Gareth Highnam %GCAT gareth862@gmail.com
of \href{http://www.bioplanet.com/}{bioplanet.com},
Filippo Spiga, %HPC
Stuart Rankin, %HPC<sjr20@cam.ac.uk>
Timothy Lanfear, %nVidia GPU vCPU speed
Neil Daeche,
Dave Twisleton,
John Andrews
and
Tristan Clark.

K20 and K40 tesla donated by \href{http://www.nvidia.com}{nVidia}.
K80 runs 
used the Darwin Supercomputer of the University
of Cambridge 
\href{http://www.hpc.cam.ac.uk/}
{High Performance Computing Service}.

\bibliographystyle{splncs} %acm} %named
\bibliography{references,gp-bibliography}

\end{document}